\documentclass{article}

\usepackage[preprint]{neurips_data_2024}

\usepackage[utf8]{inputenc} 
\usepackage[T1]{fontenc}    
\usepackage{hyperref}       
\usepackage{url}            
\usepackage{booktabs}       
\usepackage{amsfonts}       
\usepackage{nicefrac}       
\usepackage{microtype}      
\usepackage{xcolor}         
\usepackage{graphicx} 
\usepackage{array}
\usepackage{subcaption}
\usepackage{longtable}

\title{Framework for Curating Speech Datasets and Evaluating ASR Systems: A Case Study for Polish}

\author{%
  Michał Junczyk \\
  Department of Artificial Intelligence \\
  Adam Mickiewicz University, Poznań \\
  ul. Uniwersytetu Poznańskiego 4\\
  61-614 Poznań, Poland  \\
  \texttt{micjun@amu.edu.pl} \\
}

\begin{document}

\maketitle

\begin{abstract}
Speech datasets available in the public domain are often underutilized because of challenges in discoverability and interoperability. A comprehensive framework has been designed to survey, catalog, and curate available speech datasets, which allows replicable evaluation of automatic speech recognition (ASR) systems. A case study focused on the Polish language was conducted; the framework was applied to curate more than 24 datasets and evaluate 25 combinations of ASR systems and models. This research constitutes the most extensive comparison to date of both commercial and free ASR systems for the Polish language. It draws insights from 600 system-model-test set evaluations, marking a significant advancement in both scale and comprehensiveness. The results of surveys and performance comparisons are available as interactive dashboards,\footnote{\href{https://huggingface.co/spaces/amu-cai/pl-asr-leaderboard}{AMU ASR Leaderboard}} along with curated datasets\footnote{\href{https://huggingface.co/datasets/amu-cai/pl-asr-bigos-v2}{AMU BIGOS dataset}}\footnote{\href{https://huggingface.co/datasets/pelcra/pl-asr-pelcra-for-bigos}{PELCRA for BIGOS dataset}} and the open challenge call.\footnote{\href{https://poleval.pl/tasks/task3}{Polish ASR challenge}} Tools used for evaluation are open-sourced,\footnote{\href{https://github.com/goodmike31/pl-asr-bigos-tools}{AMU BIGOS Eval Tools}} facilitating replication and adaptation for other languages, as well as continuous expansion with new datasets and systems.
\end{abstract}

\section{Introduction}
\subsection{Background}
The Polish language is spoken by more than 50 million people worldwide. The number of available ASR systems and services, as well as speech data resources that support Polish, is systematically growing. However, the community lacks the resources to methodically evaluate and track progress. First, the available data assets are underutilized due to challenges such as discoverability, licensing, and interoperability. Secondly, there is no standardized ASR benchmark dataset for Poland. These issues hinder the development of new systems and applications, as reliable benchmarks and leaderboards are crucial to drive research progress and assess the suitability of ASR technologies for specific scenarios \cite{NathanLambert2023InLeaderboard}. The international ASR community has recognized the need for standardized evaluation methodologies to ensure consistent and comparative performance assessments in ASR \cite{Aksenova2021, szymanski20, gandhi22} and the ML field in general \cite{Liao2021AreLearning, Olson2017PMLB:Comparison,Northcutt2021PervasiveBenchmarks}. This calls for innovations in the management of ASR data sets and evaluation frameworks.\cite{Koo2023KEBAP:Post-processing}
 
\subsection{Research gap}
Existing data curation and ASR benchmarking methods for low-resource languages such as Polish exhibit several shortcomings:
\begin{itemize}
    \item \textbf{Data utilization:} Speech datasets are underutilized due to limited awareness or accessibility.
    \item \textbf{Data quality:} A lack of proper understanding of test sets can result in misrepresentation of current state-of-the-art performance.
    \item \textbf{Evaluation reproducibility:} Limited adoption of benchmark sets hinders the validation of the research results.
    \item \textbf{Evaluation scope:} Ecologically valid evaluation of a specific ASR application requires considering a larger number of datasets, systems, and performance metrics.
\end{itemize}

\subsection{Contributions}
\begin{enumerate}    
    \item \textbf{Curation of benchmark dataset:} A benchmark dataset was created from 24 openly available datasets to address the lack of standardized evaluation resources for Polish ASR systems. It includes robust samples from various sources of read and spontaneous speech. The dataset is openly available and actively maintained to enable systematic and comprehensive analysis.
    
    \item \textbf{Development of a benchmark framework:}  The framework supports various datasets, systems, and metrics, ensuring consistent ASR evaluation with standardized protocols.
    
    \item \textbf{Evaluation of ASR systems:} Using a curated dataset, 10 ASR systems and 25 models, both commercial and open-source, were compared. Significant variations across different systems, datasets, and speaker demographics were discovered.
    
    \item \textbf{Open sharing of resources:} All datasets, tools, and evaluation results have been made openly available to the research community. This promotes transparency, reproducibility, and collaboration, enabling other researchers to build upon the work, either by developing ASR systems for Polish based on evaluation results or applying the framework to other languages.
\end{enumerate}

\section{Methodology}

\subsection{Framework overview}
\begin{figure}
    \centering
    \includegraphics[width=0.75\linewidth]{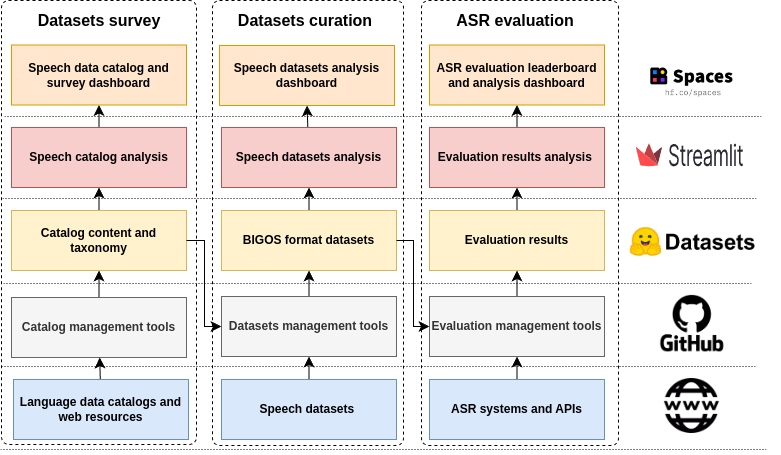}
    \caption{Architecture of data curation and ASR evaluation framework.}
    \label{fig:architecture}
\end{figure}

The devised framework for data curation and ASR benchmarking encompasses three main processes:
\begin{enumerate}
  \item \textbf{ASR speech datasets survey}
  \item \textbf{Curation of ASR benchmark dataset}
  \item \textbf{Evaluation of ASR systems}
\end{enumerate}
Figure \ref{fig:architecture} illustrates the framework architecture and the core open tools used for development. The subsequent sections provide a detailed description of the specific processes and tools.

\subsection{Survey of datasets}
A keyword-based literature review \cite{rowley2004} was used to identify and document relevant datasets. The datasets were manually analyzed and annotated. The final methodology included:
\begin{enumerate}
    \item Conducting keyword searches in relevant sources
    \item Manually analyzing and annotating documentation
    \item Cross-checking multiple sources for consistency and accuracy
    \item Validating and analyzing downloadable datasets
    \item Analyzing metadata to derive insights on Polish ASR speech datasets
    \item Making the catalog and insights publicly available
\end{enumerate}

The survey sources include language data repositories, scientific community platforms, and public domain documentation. The attributes considered include creator, funding, license, publication date, quality assurance, and content characteristics such as the format of the audio file and the number of speakers \cite{Junczyk2024ADatasets}. Resulting catalog and survey insights are shared on GitHub\footnote{\href{https://github.com/goodmike31/pl-asr-speech-data-survey/}{Polish ASR speech data survey -- GitHub}} and Hugging Face.\footnote{\href{https://huggingface.co/spaces/amu-cai/pl-asr-survey}{Polish ASR survey -- Hugging Face}}

\subsection{Dataset curation}
\subsubsection{Design considerations}
A curated benchmark dataset for Polish ASR systems is intended to have the following features: 
\begin{itemize}
    \item \textbf{Task-appropriate: } Relevant and practical for the intended ASR task.
    \item \textbf{Accessible: }Available online under a license that allows the free use and creation of derivative works.
    \item \textbf{Discoverable:} Easy to find and acquire (without time-consuming registration or other access barriers).
    \item \textbf{Diverse and challenging:} Containing various examples to test the adaptability of the model, as well as complex cases to encourage community participation and minimize the risk of benchmark saturation.
    \item \textbf{Annotated}: With metadata about speakers and recordings allowing nuanced analysis and interpretation of the results.
    \item \textbf{Optimally sized:} Large enough to be representative, but manageable to download and explore.
    \item \textbf{Clean yet realistic: }Free of major errors, but noisy enough to represent the complexity of the real world.
    \item \textbf{Well-documented: } Provided with documentation that is understandable to users without technical skills.
    \item \textbf{Well-explained:} Provided with evaluation baselines and how-to-use script examples. 
\end{itemize}

\subsubsection{Leveraging speech data catalog for sourcing open data sets}
The Polish ASR speech dataset catalog \cite{Junczyk2023PolishCatalog} was used to select datasets for curation based on following criteria:
\begin{itemize}
\item Datasets are available online under a license allowing free use for non-commercial purposes.
\item Transcriptions are aligned with the recordings.
\item Recording sampling rate is at least 8 kHz.
\item Audio files are encoded using at least 16 bits per sample.
\end{itemize}

24 datasets were selected for curation as \textit{BIGOS\footnote{The Polish word \textit{bigos} is the name of a cabbage-based stew.} (Benchmark Intended Grouping of Open Speech)} benchmark dataset:

\begin{itemize}
\item \textbf{The Common Voice data set} \emph{(mozilla-common\_voice\_15-23)} is a multilingual resource \cite{Ardila2019CommonCorpus} covering over 60 languages and many underrepresented groups. Available under CC-0 license.

\item \textbf{The Multilingual LibriSpeech (MLS) data set} \emph{(fair-mls-20)} is a large multilingual corpus made by Facebook AI Research (FAIR) \cite{Pratap2020MLS:Research}. Derived from audiobooks, it covers eight languages, with 44,000 hours of English and 6,000 hours for other languages. The Polish data includes 137 hours from 25 books by 16 speakers. Available under CC-BY license.

\item \textbf{The Clarin Studio data set} \emph{(clarin-pjatk-studio-15)} by CLARIN-PL includes 13,802 short utterances (56 hours) from 554 sessions by 317 speakers. Each session has 20-31 audio files, all recorded in a studio for clear audio. Available under CC-BY-SA license.

\item \textbf{The Clarin Mobile data set} \emph{(clarin-pjatk-mobile-15)} is a Polish speech corpus of read speech recorded on a telephone. It includes many speakers reading several dozen sentences and words with rare phonemes. Available under CC-BY-SA license.

\item \textbf{The Jerzy Sas PWR data sets} (Politechnika Wrocławska) comprise three legacy sets of recordings available in the public domain:
\begin{itemize}
\item Male speaker speech set \textit{(pwr-maleset-unk)} – single male speaker recordings.
\item Utterances containing short words (\textit{pwr-shortwords-unk}) – single-phoneme conjunctions and prepositions likely to be misrecognized.
\item Spoken commands as very important utterances (VIUs) \textit{(pwr-viu-unk)} – editor control commands and  domain-specific utterances.
\end{itemize}

\item \textbf{The M-AI Labs Speech corpus} \emph{(mailabs-19)} created from audiobooks as \textit{MLS}.  Intended for training speech recognition and synthesis systems in nine languages, with nearly a thousand hours of audio, including 53.5 hours for Polish. Available under proprietary license.

\item \textbf{The AZON Read and Spontaneous Speech data sets} \emph{(pwr-azon\_spont-20, pwr-azon\_read-20)} contain recordings from academic staff in the physical chemistry domain, including both supervised readings and unsupervised spontaneous recordings such as interviews and presentations. Available under a CC-BY-SA license.\footnote{\href{https://zasobynauki.pl/zasoby/korpus-nagran-probek-mowy-do-celow-budowy-modeli-akustycznych-dla-automatycznego-rozpoznawania-mowy,53293/}{AZON dataset homepage}}

\item \textbf{Google FLEURS} \emph{(google-fleurs-22)} is a parallel speech benchmark data set in 102 languages, based on the FLoRes-101 machine translation benchmark \cite{Conneau2022FLEURS:Speech}. Hosted on Hugging Face\footnote{\href{https://huggingface.co/data sets/google/fleurs}{FLEURS dataset homepage}} and available under a CC-BY license.

\item \textbf{PolyAI Minds14} (\emph{polyai-minds14-21}) is a dataset for training and evaluating intent recognition systems using spoken data. Covers spoken samples in the commercial e-banking domain in 14 language variations \cite{Gerz2021MultilingualData}. Hosted on Hugging Face\footnote{\href{https://huggingface.co/data sets/PolyAI/minds14}{Minds14 dataset homepage}} and available under a CC-BY license.

\item \textbf{PolEval 22 Diabiz sample} (\emph{ul-diabiz\_poleval-22)} was used for a punctuation restoration task in the 2022 PolEval competition. It is a subset of the \textit{DiaBiz homepage}\footnote{\href{http://docs.pelcra.pl/doku.php?id=diabiz}{Diabiz}} dialog corpus of phone-based customer--agent interactions by the PELCRA group of the University of Łódź. Available publicly under CC-BY-SA-NC-ND and curated with the consent of the author.

\item \textbf{SpokesMix}\footnote{\href{http://docs.pelcra.pl/doku.php?id=spoken_offline_corpora}{SpokesMix dataset homepage}} is a corpus of conversational Polish by the PELCRA group \cite{spokesmix2018}. It includes speech recordings and word-by-word transcriptions with non-speech events. Available under the CC-BY-NC-ND license and curated for ASR benchmarking purposes with permission of the author.

\item \textbf{SpokesBiz}\footnote{\href{http://docs.pelcra.pl/doku.php?id=spokesbiz\%22}{SpokesBiz dataset homepage}} is a corpus of conversational Polish from the CLARIN-BIZ project, featuring over 650 hours of recordings from nearly 600 speakers \cite{pezik2023spokesbiz}. Transcriptions are diarized and manually annotated. Includes eight diverse subsets, e.g. biographical interviews, job interviews, podcasts, and student presentations. Available under the CC-BY-NC-ND license and curated for ASR benchmarking purposes with the author's permission.
\end{itemize}

Datasheets describing the content of curated datasets can be found in Appendices \ref{app:dataset_stats}, \ref{app:dataset_features_speech_type}, \ref{app:dataset_features_license} and \ref{app:dataset_features_env_device}, as well as Hugging Face.\footnote{\href{https://huggingface.co/spaces/amu-cai/amu-bigos-data-dash}{AMU BIGOS Dataset Dashboard}}

\subsubsection{Curation process}
\begin{enumerate}
    \item \textbf{Dataset structure curation:}
  \begin{itemize}
    \item Downloading and manually inspecting format and contents
    \item Creating train/dev/test splits if not available
    \item Assigning standard IDs to speakers and files
  \end{itemize}
    \item \textbf{Audio file curation:}
  \begin{itemize}
    \item Removal of invalid audio files
    \item Unifying audio format to WAV 16 bits/16 kHz
    \item Normalizing audio amplitude to -3 dBFS
    \item Splitting long audio files into shorter segments based on time-alignment annotations
  \end{itemize}
    \item \textbf{Text files (transcripts and metadata) curation:}
  \begin{itemize}
    \item Converting text encoding to UTF8 
    \item Extracting original transcription and removing redundant characters
    \item Extracting and unifying metadata contents
    \item Generating metadata from text and audio content
    \item Saving in the standard tabular format 
    \end{itemize}
\item \textbf{Dataset distribution}
    \begin{itemize}
    \item Uploading to the HF dataset hub
    \item Referencing the original license in the README file
    \end{itemize}
\end{enumerate}

The resulting \textit{BIGOS utterance data object} with a description of the standard metadata fields is available in Table \ref{tab:utterance_data_object} in the Appendix.

\subsection{ASR evaluation}
\subsubsection{System design considerations}
Established tools and platforms were used where possible. Table \ref{tab:bench_system_design_considerations} provides an overview of the main design considerations.

\begin{table}[htbp]
\centering
\caption{Design considerations for ASR evaluation system.}
\label{tab:bench_system_design_considerations}
\begin{tabular}{>{\raggedright} p{3cm}  >{\raggedright\arraybackslash}p{10cm} }
\toprule

\textbf{Aspect} & \textbf{Considerations} \\
\midrule
Metrics& Support for well-established metrics.\\ 
Extensibility& Straightforward integration of new datasets, normalization methods, metrics and new ASR systems.\\ 
\midrule
Availability& Publicly accessible and intuitive presentation of results.\\ 
\midrule
Comprehensiveness& Performance analysis across scenarios, system params and user groups.\\ 
\bottomrule
\end{tabular}

\end{table}

\subsubsection{Overview of the evaluation process}
In total 25 models of 7 ASR systems were evaluated: Google STT, Azure STT, Whisper, AssemblyAI,  NeMo, MMS and Wav2Vec. The complete list is presented in Table \ref{tab:asr_systems_evaluated_names}. Currently, 5 evaluation metrics are supported: SER, WER, MER, WIL, and CER \cite{Morris2004FromRecognition}. The methods for normalizing references and hypotheses are listed in Appendix \ref{tab:normalization_methods}. Python scripts used for the evaluation are available on GitHub.\footnote{\href{https://github.com/goodmike31/pl-asr-bigos-tools}{BIGOS ASR evaluation tools}}

\begin{figure}
    \centering
    \includegraphics[width=0.8\linewidth]{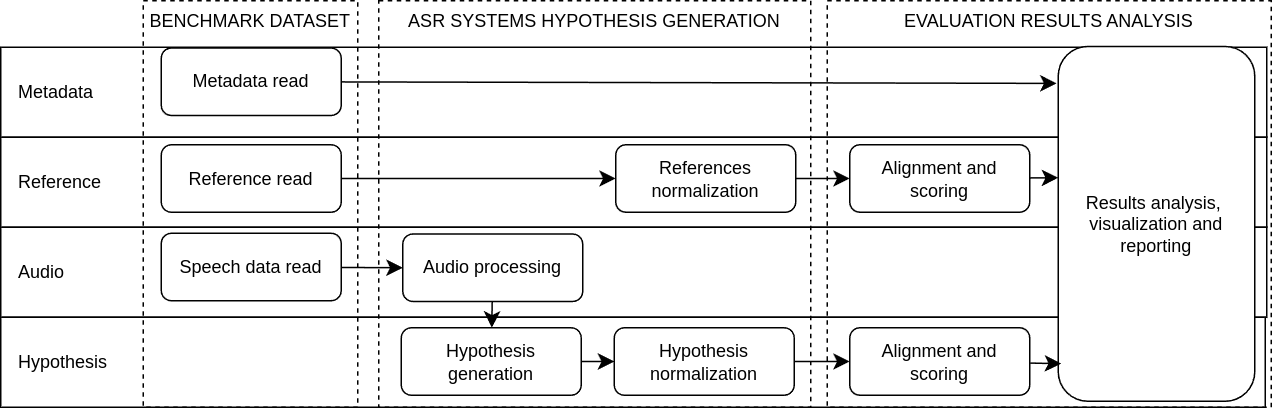}
    \caption{ASR evaluation process data flow}
    \label{fig:eval_data_flow}
\end{figure}

\section{Evaluation results}
The developed framework supports the following evaluation scenarios.
\begin{table}[bp]
  \centering
    \caption{Evaluation scenarios and their corresponding analysis dimensions and metrics}
    \label{tab:evaluation_scenarios}
  \begin{tabular}{l p{8cm}}
    \toprule
    \textbf{Scenario codename} & \textbf{Analysis dimensions of accuracy}\\ \midrule
    Per Systems & Specific system--model variants across all dataset subsets.\\ \midrule
    Per Subset & System--model variants across specific subsets of test data.\\ \midrule
    Per System Type & Free and commercial systems, including best and worse performing ones.\\ \midrule
    Per Model Size & System--model variants with known model sizes.\\ \midrule
    Per Audio Duration & Best performing systems across audio duration ranges.\\ \midrule
    Per Speaking Rate & Best performing systems across speech rate ranges.\\ \midrule
    Per Speaker Age Group & System--model variants across test data with age info.\\ \midrule
    Per Speaker Gender & System--model variants across test data with gender info.\\ 
    \bottomrule
  \end{tabular}
\end{table}
The results of selected scenarios are analyzed in the subsequent sections. Additional results are available in Appendix \ref{app:evaluation_results}. All and more detailed results can be accessed through the public dashboard.\footnote{\href{https://huggingface.co/spaces/amu-cai/pl-asr-leaderboard}{AMU Polish ASR Leaderboard}} Dashboard users can display the evaluation results for a specific scenario and choose between various datasets, systems, metrics, normalization techniques, and diagram types.

\subsection{Impact of normalization on error rates}
Table \ref{tab:normalization_impact_bigos} shows the specific and average reduction of error rates in percentage points depending on the applied normalization method.

\begin{table}[htbp]
\centering
\caption{Reduction of error rates caused by normalization of references and hypothesis for BIGOS dataset.}
\label{tab:normalization_impact_bigos}
\begin{tabular}{l l l l l l}
\toprule
\textbf{Method}& \textbf{SER [p.p.]}& \textbf{WER [p.p.]}& \textbf{MER [p.p.]}& \textbf{CER [p.p.]}& \textbf{Average [p.p.]}\\ \midrule 
blanks& -1.53& 0& 0& -0.85& -0.6\\ \midrule 
lowercase& -2.71& -6.37& -6.59& -1.47& -4.28\\ \midrule 
punctuation& -1.84& -8.11& -8.47& -1.77& -5.05\\ \midrule 
all& -25.02& -15.52& -16.15& -4.19& -15.22\\ 
\bottomrule
\end{tabular}
\end{table}

\subsection{Overall accuracy of available ASR systems and models}
\begin{figure}
    \centering
    \includegraphics[width=0.75\linewidth]{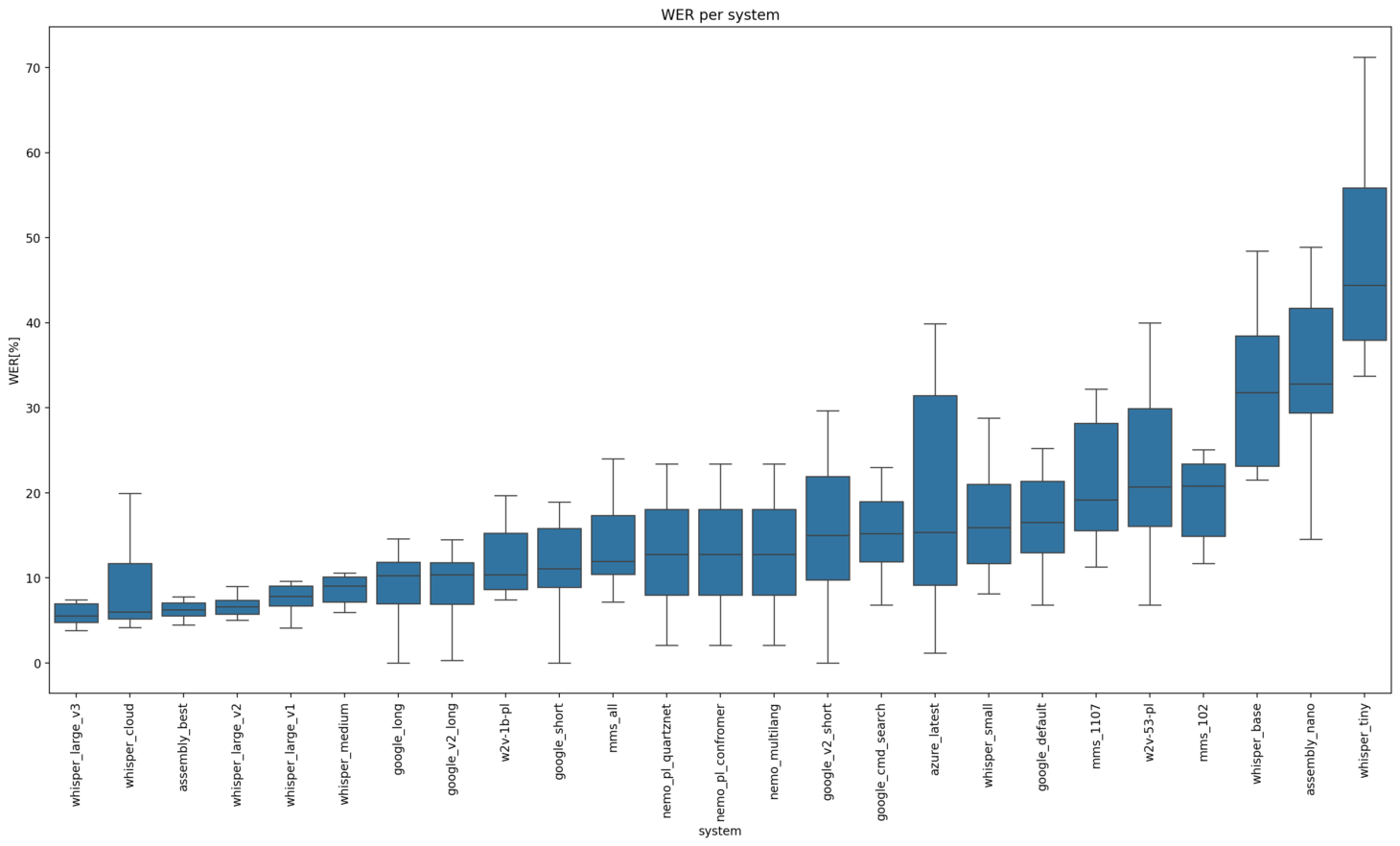}
    \caption{Box plot of WER for systems evaluated on the BIGOS dataset.}
    \label{fig:wer_systems}
\end{figure}

Figure \ref{fig:wer_systems} shows the WER box plot for the systems evaluated using the BIGOS dataset. The 3 best ASR models in terms of accuracy are \emph{Whisper Large V3, Whisper Cloud} and \emph{Assembly AI best}. The results of the evaluation using the PELCRA dataset are available in the Polish ASR leaderboard \footnote{\href{https://huggingface.co/spaces/amu-cai/pl-asr-leaderboard}{AMU ASR Leaderboard}}

\subsection{Comparison of accuracy of commercial and freely available ASR systems }
Table \ref{tab:free_vs_commercial_bigos} compares the Word Error Rate (WER) of commercial and free ASR systems. Commercial systems achieve better  median, mean and minimal error rates in the BIGOS and PELCRA datasets by approximately 2.5 p.p. and 3.5 p.p., respectively. Furthermore, commercial and free systems show better accuracy for read speech than conversational speech by approximately 17 and 18.5 p.p., respectively. 
\begin{table}[htbp]
\centering
\caption{WER statistics for freely available and commercial ASR systems}
\label{tab:free_vs_commercial_bigos}

\begin{tabular}{c c c r r r r }
\bottomrule
 \textbf{Dataset}&\textbf{Speech}&\textbf{Systems}& \textbf{Med. WER}& \textbf{Mean WER}& \textbf{Std. WER}& \textbf{Min. WER}\\
 \midrule
 BIGOS&read&comm.& 12.96 & 17.26 & 24.98 & 0 \\ \midrule 
 BIGOS&read&free & 14.57 & 19.76 & 17.36 & 2.1 \\ \midrule
 PELCRA&spont.& comm.& 29.88& 31.29& 14.69& 5.27\\\midrule
 PELCRA&spont.& free& 33.58& 35.67& 17.36& 8.74\\
 \bottomrule
\end{tabular}

\end{table}

\subsection{Accuracy as a function of model size}
Figure \ref{fig:wer_model_size} shows that as model size increases, WER decreases, indicating better performance. This trend holds for models of the same type, e.g., \emph{whisper} models. There are noticeable accuracy differences in models of the same size trained on different data, such as MMS. Finally, \emph{Nemo} models perform on par with much larger \emph{wav2vec2} models. 

\subsection{Accuracy as a function of speech rate}
\begin{figure}[htbp]
    \centering
    \begin{subfigure}[b]{0.45\linewidth}
        \centering
            \includegraphics[width=\linewidth]{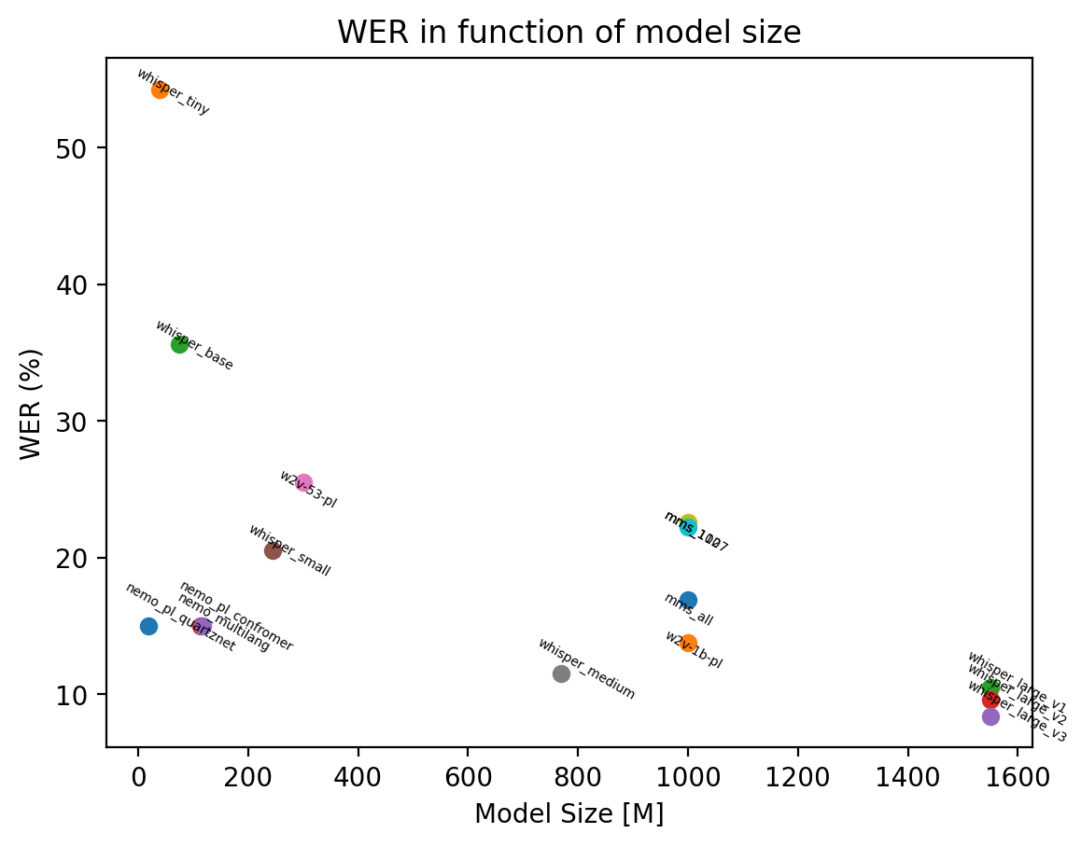}
            \caption{Accuracy as a function of model size.}
            \label{fig:wer_model_size}
    \end{subfigure}
    \hfill
    \begin{subfigure}[b]{0.45\linewidth}
        \centering
        \includegraphics[width=\linewidth]{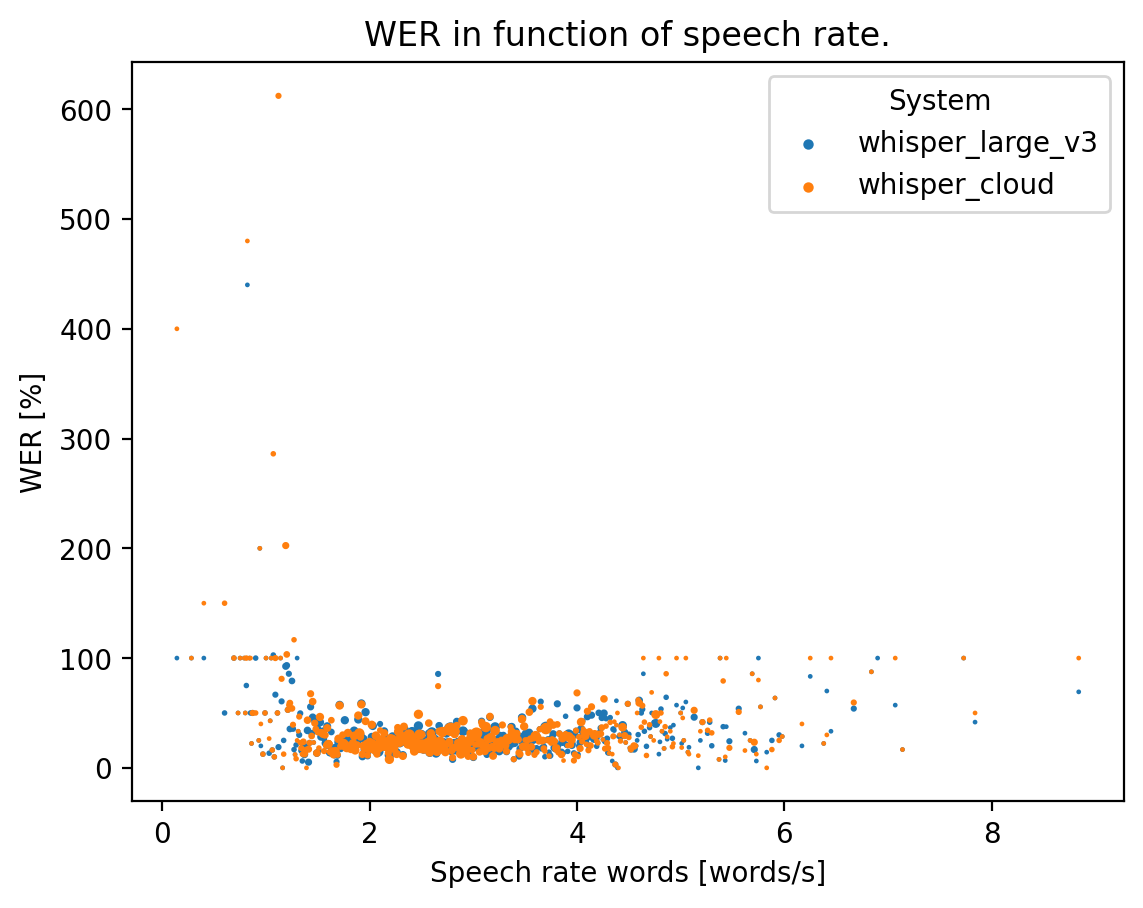}
        \caption{Accuracy as a function of speech rate range.}
        \label{fig:wer_speech_rate_pelcra}
    \end{subfigure}
    \caption{Example evaluation results available on the Polish ASR quality dashboard.}
    \label{fig:evaluation_dashboards_examples}
\end{figure}

Figure \ref{fig:wer_speech_rate_pelcra} illustrates the correlation between WER and speech rate, which is measured as the mean number of words uttered per second.

\section{Discussion}
\subsection{Analysis of findings}
\subsubsection{Impact of normalization}
Normalization techniques resulted in significant reductions in error rates for all types of metrics (SER, WER, MER, CER). Applying all methods reduced WER by 16.07 p.p. for the PELCRA dataset and 15.52 p.p. for the BIGOS dataset, highlighting the sensitivity of lexical metrics to spelling and formatting variations.

\subsubsection{Determining the best systems among free and commercial}
Conversational speech (PELCRA) has higher error rates due to its spontaneous nature, with more variability in style, speed, and pauses. Read speech (BIGOS) is more structured and consistent, resulting in lower WERs.

\subsubsection{Impact of model size on accuracy}
\begin{itemize}
    \item \verb|whisper_large v2|, \verb|whisper_large|, and \verb|whisper_large v3| show the best performance with the lowest WERs and the largest model sizes.
    \item \verb|whisper_tiny| is the second smallest model and has the highest WER among all evaluated.
    \item \verb|nemo_pl_quartznet| and \verb|nemo_pl_multilang| are relatively small models with reasonably low WERs, indicating that they are efficient given their size.
\end{itemize}

\subsubsection{Impact of speech rate on accuracy (WER)}
\begin{itemize}

    \item Both \verb|whisper_large_v3| and \verb|whisper_cloud| perform similarly across speech rates. For rates between 1.5 and 5, most WERs are below 30\%. Severe errors occur at lower rates, while higher rates increase WER, indicating limited robustness for faster speech. Outliers suggest challenging scenarios or truncated audio/transcriptions.
\end{itemize}

\subsection{Implications}
The developed data curation and evaluation framework offers the following benefits for the research community:
\begin{itemize}
\item Establishes a consistent framework for evaluating Polish ASR systems, enhancing reproducibility.
\item Facilitates better use of datasets, promoting focused research.
\item Encourages data sharing and collaboration, improving resources and progress.
\item Identifies gaps, such as the need for detailed metadata and semantic metrics, guiding future studies.
\end{itemize}

Advantages for industry include:
\begin{itemize}
  \item Informs public about strengths and weaknesses of available ASR system.
  \item Proposes a standard evaluation procedure to increase evaluation efficiency.
  \item Showcases the importance of normalization and utilization of metadata for analysis.
  \item Provides incentive to companies to showcase superior performance on a public benchmark for marketing purposes.
\end{itemize}

\subsection{Limitations and challenges}
Future research should include manual transcriptions and annotations to assess the quality of test data \cite{Koo2024TowardGuideline}. Investigating manual annotation of recognition errors to determine the criticality of the error \cite{Wirth2022AutomaticAnalysis}, and automating the classification and correction of erroneous references are other directions to explore. Integrating semantically informed metrics could provide additional insight into accuracy\cite{Stokke2023SemanticDistance, Roy2021Semantic-WER:Usability}. Robustness and bias measurements could be improved by augmenting existing or collecting new recordings representing various usage conditions and Polish speakers demographics.\cite{Aksenova2021,Aksenova2022AccentedData}

\section{Conclusion}
The research establishes a framework for evaluating ASR systems. It addresses the issue of limited dataset usage for Polish benchmarking by offering a curated benchmark set derived from 24 publicly available datasets identified in an extensive survey. The evaluation of 7 ASR systems and 25 models revealed notable performance differences between service types, model sizes, and speech types. The study also highlighted potential problems with the test set content that require further examination. This work improves reproducibility and directs future ASR advancements by providing public access to data catalogs, curated datasets, evaluation tools, and dashboards with benchmarking results.

\section{References}
\bibliography{references}

\section{Appendices}
Provide additional data, tools' documentation, and other supplementary materials that are relevant but not central to the article's narrative.

\section*{Checklist}
\begin{enumerate}

\item For all authors...
\begin{enumerate}
  \item Do the main claims made in the abstract and introduction accurately reflect the paper's contributions and scope?
    \answerYes{Abstract and introduction explicitely describes contributions: Survey of datasets, metholodogy thereof, curated evaluation datasets process and outcomes, system for ASR evaluation, interactive dashboard with benchmark results.}
  \item Did you describe the limitations of your work?
    \answerYes{Limitations include limited representation of Polish speakers, lack of manual transcription verification and unification, limited scope of transcription normalization, lack of support for embedding based metrics, lack of manual analysis of ASR errors, limited availability of recordings with speaker metadata.}
  \item Did you discuss any potential negative societal impacts of your work?
    \answerYes{In the limitations section, it is mentioned that the evaluation datasets do not encompass all Polish users or the various conditions under which ASR systems are used. However, the results presented can guide the selection of the best-performing ASR systems for use-cases similar to those in the BIGOS evaluation dataset. For new and particularly high-risk scenarios, such as the medical field or specific demographic group, an independent evaluation on a representative dataset is necessary to accurately assess performance and ensure safe, unbiased operation.}
  \item Have you read the ethics review guidelines and ensured that your paper conforms to them?
    \answerYes{Presented work follows the ethical guidelines. No PII or protected information about individuals is revealed. The author obtained consent to use datasets for evaluation dataset curation and evaluation, either directly or based on licensing terms. Research did not include experiments involving human subjects.}
\end{enumerate}

\item If you are including theoretical results...
\begin{enumerate}
  \item Did you state the full set of assumptions of all theoretical results?
    \answerNA{}
	\item Did you include complete proofs of all theoretical results?
    \answerNA{}{}
\end{enumerate}

\item If you ran experiments (e.g. for benchmarks)...
\begin{enumerate}
  \item Did you include the code, data, and instructions needed to reproduce the main experimental results (either in the supplemental material or as a URL)?
    \answerYes{Code, data and instructions how to reproduce results are available on respective publicly available repositories on Hugging Face and GitHub platforms.}
  \item Did you specify all the training details (e.g., data splits, hyperparameters, how they were chosen)?
    \answerNA{}
	\item Did you report error bars (e.g., with respect to the random seed after running experiments multiple times)?
    \answerNA{}
	\item Did you include the total amount of compute and the type of resources used (e.g., type of GPUs, internal cluster, or cloud provider)?
    \answerNA{}
\end{enumerate}

\item If you are using existing assets (e.g., code, data, models) or curating/releasing new assets...
\begin{enumerate}
  \item If your work uses existing assets, did you cite the creators?
    \answerYes{Yes, all authors of existings assets were cited both in submitted article and repositories with curated assets.}
  \item Did you mention the license of the assets?
    \answerYes{Yes, license types are mentioned in the respective tables describing source datasets, as well as on repositories hosting curated assets.}
  \item Did you include any new assets either in the supplemental material or as a URL?
    \answerYes{Yes, links to meta-corpora resulting from curation of existing assests were provided.}
  \item Did you discuss whether and how consent was obtained from people whose data you're using/curating?
    \answerYes{Yes, the consent from the author of PELCRA corpora to curate dataset for open competition and benchmarking purposes is mentioned.}
  \item Did you discuss whether the data you are using/curating contains personally identifiable information or offensive content?
    \answerYes{Yes, the lack of PII is mentioned, however the inspection if datasets contain potentially offensive content was not performed.}
\end{enumerate}

\item If you used crowdsourcing or conducted research with human subjects...
\begin{enumerate}
  \item Did you include the full text of instructions given to participants and screenshots, if applicable?
    \answerNA{}
  \item Did you describe any potential participant risks, with links to Institutional Review Board (IRB) approvals, if applicable?
    \answerNA{}
  \item Did you include the estimated hourly wage paid to participants and the total amount spent on participant compensation?
    \answerNA{}
\end{enumerate}

\end{enumerate}


\appendix

\section{Additional information required by organizers}
In the Appendix, we provide additional information. This section will often be part of the supplemental material. Please see the call on the NeurIPS website for links to additional guides on dataset publication.

Submission introducing new datasets must include the following in the supplementary materials:
\begin{enumerate}
  \item Dataset documentation and intended uses. Recommended documentation frameworks include datasheets for datasets, dataset nutrition labels, data statements for NLP, and accountability frameworks.
  \item URL to website/platform where the dataset/benchmark can be viewed and downloaded by the reviewers.
  \item URL to Croissant metadata record documenting the dataset/benchmark available for viewing and downloading by the reviewers. You can create your Croissant metadata using e.g. the Python library available here: https://github.com/mlcommons/croissant
  \item Author statement that they bear all responsibility in case of violation of rights, etc., and confirmation of the data license.
  \item Hosting, licensing, and maintenance plan. The choice of hosting platform is yours, as long as you ensure access to the data (possibly through a curated interface) and will provide the necessary maintenance.
\end{enumerate}

\section{Additional information relevant to submitted article}

\subsection{Dataset splits details}\label{app:dataset_splits}
Tables \ref{tab:subsets_partitioning_bigos} and \ref{tab:subsets_partitioning_pelcra} present logic of data splits applied during curation for BIGOS and PELCRA datasets, respectively.
\begin{table}[htbp]
    \caption{Metadata and partitioning of source datasets -- BIGOS dataset}
    \label{tab:subsets_partitioning_bigos}
    \centering
    \begin{tabular}{l l l l}
        \toprule
        \textbf{Subset} & \textbf{Original partitioning} & \textbf{BIGOS split process} & \textbf{Entity for BIGOS split} \\
        \midrule
        google-fleurs-22              & train, test, dev      & original splits preserved & N/A \\
        \midrule
        polyai-minds14-21             & none                  & pseudorandom              & audio file id \\
        \midrule
        pjatk-clarin\_mobile-15       & none                  & pseudorandom              & session (speaker id) \\
        \midrule
        pjatk-clarin\_studio-15       & none                  & pseudorandom              & session (speaker id) \\
        \midrule
        pwr-azon\_read-20             & none                  & pseudorandom              & session (speaker id) \\
        \midrule
        pwr-azon\_spont-20            & none                  & pseudorandom              & session (speaker id) \\
        \midrule
        fair-mls-20                   & train, test, dev      & original splits preserved & N/A \\
        \midrule
        mozilla-cv15-23               & train, test, dev      & original splits preserved & N/A \\
        \midrule
        mailabs-corpus\_librivox-19   & none                  & pseudorandom              & audio file id \\
        \midrule
        pwr-maleset-unk               & none                  & pseudorandom              & audio file id \\
        \midrule
        pwr-shortwords-unk            & none                  & pseudorandom              & audio file id \\
        \midrule
        pwr-viu-unk                   & none                  & pseudorandom              & audio file id \\
        \bottomrule
    \end{tabular}
\end{table}

\begin{table}[htbp]
    \caption{Metadata and partitioning of source datasets -- PELCRA Dataset}
    \label{tab:subsets_partitioning_pelcra}
    \centering
    \begin{tabular}{l l l l}
        \toprule
        \textbf{Subset} & \textbf{Original partitioning} & \textbf{BIGOS split process} & \textbf{Entity for BIGOS split} \\
        \midrule
        ul-diabiz\_poleval-22       & train, test, dev & original splits preserved & N/A \\
        \midrule
        ul-spokes\_biz\_bio-23      & none             & pseudorandom              & recording id \\
        \midrule
        ul-spokes\_biz\_int-23      & none             & pseudorandom              & recording id \\
        \midrule
        ul-spokes\_biz\_luz-23      & none             & pseudorandom              & recording id \\
        \midrule
        ul-spokes\_biz\_pod-23      & none             & pseudorandom              & recording id \\
        \midrule
        ul-spokes\_biz\_pres-23     & none             & pseudorandom              & recording id \\
        \midrule
        ul-spokes\_biz\_vc-23       & none             & pseudorandom              & recording id \\
        \midrule
        ul-spokes\_biz\_vc2-23      & none             & pseudorandom              & recording id \\
        \midrule
        ul-spokes\_biz\_wyw-23      & none             & pseudorandom              & recording id \\
        \midrule
        ul-spokes\_mix\_emo-18      & none             & pseudorandom              & recording id \\
        \midrule
        ul-spokes\_mix\_luz-18      & none             & pseudorandom              & recording id \\
        \midrule
        ul-spokes\_mix\_parl-18     & none             & pseudorandom              & recording id \\
        \bottomrule
    \end{tabular}
\end{table}

\subsection{Dataset splits details}\label{app:bigos_data_object_attributes}
Table \ref{tab:utterance_data_object} presents metadata fields associated with each individual data item in BIGOS datasets.
\begin{table}[htbp]
    \caption{Attributes in the BIGOS utterance data object}
    \label{tab:utterance_data_object}
    \centering
    \begin{tabular}{>{\raggedright}p{3cm} >{\raggedright\arraybackslash}p{10cm}}
        \toprule
        \textbf{Field name} & \textbf{Description} \\
        \midrule
        audioname & Standardized unique identifier for each audio recording in the dataset e.g. \\
        \midrule
        split & Indicates the dataset split the recording belongs (e.g., train, test, validation). \\
        \midrule
        dataset & Source dataset identifier \\
        \midrule
        ref\_orig & The original transcript associated with the audio recording. \\
        \midrule
        ref\_spoken & Transcription in the spoken domain format. \\
        \midrule
        ref\_written & Transcription in the written domain format. \\
        \midrule
        audio & Object for storing audio data in HF datasets format. \\
        \midrule
        sampling\_rate & The sampling rate of the audio recording in the dataset. Can be the same as the original or adjusted for standardization. \\
        \midrule
        samplingrate\_orig & The original sampling rate of the audio recording. \\
        \midrule
        speaker\_id & A unique identifier of the speaker in the recording. \\
        \midrule
        audiopath\_bigos & The relative path to the audio file from distributed data archive. \\
        \midrule
        audiopath\_local & The absolute path to the extracted audio file, typically in the default hf datasets cache directory. \\
        \midrule
        audio\_duration\_samples & Recording duration in samples. \\
        \midrule
        audio\_duration\_seconds & Recording duration in seconds. \\
        \midrule
        speaker\_gender & Information about the speaker's gender in the CommonVoice format. If not available, it is indicated as N/A (Not Available). \\
        \midrule
        speaker\_age & Information about the speaker's age in CommonVoice format. If not available, it is indicated as N/A (Not Available). \\
        \midrule
        speech\_rate\_words & Speech rate expressed in words per seconds. \\
        \midrule
        speech\_rate\_chars & Speech rate expressed in characters per seconds. \\
        \midrule
        utterance\_length\_words & Length of the utterance in words. \\
        \midrule
        utterance\_length\_chars & Length of the utterance in characters. \\
        \bottomrule
    \end{tabular}
\end{table}

\subsection{Dataset contents details}\label{app:dataset_features_license}
Tables \ref{tab:info_license_langs_bigos} and \ref{tab:info_license_langs_pelcra} present information on licensing and language coverage for BIGOS and PELCRA datasets, respectively.

\begin{table}[htbp]
    \caption{BIGOS V2 dataset subset license and language coverage.}
    \label{tab:info_license_langs_bigos}
    \centering
    \begin{tabular}{c c c c}
        \toprule
        \textbf{Dataset} & \textbf{Codename} & \textbf{License} & \textbf{Languages} \\
        \midrule
        Clarin Studio                   & pjatk-clarin\_studio-15   & CC-BY         & monolingual  \\
        \midrule
        Clarin Mobile                   & pjatk-clarin\_mobile-15   & CC-BY         & monolingual  \\
        \midrule
        Munich AI Labs LibriVox         & mailabs-corpus\_librivox-19 & Proprietary  & multilingual \\
        \midrule
        Mozilla Common Voice            & mozilla-common\_voice\_15-23 & CC-0         & multilingual \\
        \midrule
        Multilingual Librispeech        & fair-mls-20               & CC-BY         & multilingual \\
        \midrule
        Azon Read                       & pwr-azon\_read-20         & CC-BY-SA      & monolingual  \\
        \midrule
        Azon Spontaneous                & pwr-azon\_spont-20        & CC-BY-SA      & monolingual  \\
        \midrule
        PWR Male Set                    & pwr-maleset-unk           & Public domain & monolingual  \\
        \midrule
        PWR Short Words                 & pwr-shortwords-unk        & Public domain & monolingual  \\
        \midrule
        PWR Very Important Utterances   & pwr-viu-unk               & Public domain & monolingual  \\
        \midrule
        Google FLEURS                   & google-fleurs-22          & CC-BY         & multilingual \\
        \midrule
        PolyAI Minds14                  & polyai-minds14-21         & CC-BY         & multilingual \\
        \bottomrule
    \end{tabular}
\end{table}

\begin{table}[htbp]
    \caption{PELCRA for BIGOS dataset subset license and language coverage.}
    \label{tab:info_license_langs_pelcra}
    \centering
    \begin{tabular}{c c c c}
        \toprule
        \textbf{Dataset} & \textbf{Codename} & \textbf{License} & \textbf{Languages} \\
        \midrule
        DiaBiz ASR PolEval 22      & ul-diabiz\_poleval-22      & Public domain     & monolingual \\
        \midrule
        SpokesBiz CBIZ\_BIO        & ul-spokes\_biz\_bio-23     & CC-BY-NC-ND       & monolingual \\
        \midrule
        SpokesBiz CBIZ\_INT        & ul-spokes\_biz\_int-23     & CC-BY-NC-ND       & monolingual \\
        \midrule
        SpokesBiz CBIZ\_LUZ        & ul-spokes\_biz\_luz-23     & CC-BY-NC-ND       & monolingual \\
        \midrule
        SpokesBiz CBIZ\_POD        & ul-spokes\_biz\_pod-23     & CC-BY-NC-ND       & monolingual \\
        \midrule
        SpokesBiz CBIZ\_PRES       & ul-spokes\_biz\_pres-23    & CC-BY-NC-ND       & monolingual \\
        \midrule
        SpokesBiz CBIZ\_VC         & ul-spokes\_biz\_vc-23      & CC-BY-NC-ND       & monolingual \\
        \midrule
        SpokesBiz CBIZ\_VC2        & ul-spokes\_biz\_vc2-23     & CC-BY-NC-ND       & monolingual \\
        \midrule
        SpokesBiz CBIZ\_WYW        & ul-spokes\_biz\_wyw-23     & CC-BY-NC-ND       & monolingual \\
        \midrule
        SpokesMix PELCRA\_EMO      & ul-spokes\_mix\_emo-18     & CC-BY             & monolingual \\
        \midrule
        SpokesMix PELCRA\_LUZ      & ul-spokes\_mix\_luz-18     & CC-BY             & monolingual \\
        \midrule
        SpokesMix PELCRA\_PARL     & ul-spokes\_mix\_parl-18    & CC-BY             & monolingual \\
        \bottomrule
    \end{tabular}
\end{table}
\subsection{Dataset contents details}\label{app:dataset_features_speech_type}
Tables \ref{tab:domains_speech_type_bigos} and \ref{tab:domains_speech_type_pelcra} present information on domains, speech, and interaction types for BIGOS and PELCRA datasets, respectively.

\begin{table}[htbp]
    \caption{BIGOS V2 dataset subset domains and speech types.}
    \label{tab:domains_speech_type_bigos}
    \centering
    \begin{tabular}{c c c c}
        \toprule
        \textbf{Codename} & \textbf{Domain} & \textbf{Speech type} & \textbf{Interaction type} \\
        \midrule
        pjatk-clarin\_studio-15     & open domain   & read        & monolog \\
        \midrule
        pjatk-clarin\_mobile-15     & open domain   & read        & monolog \\
        \midrule
        mailabs-corpus\_librivox-19 & audiobook     & read        & monolog \\
        \midrule
        mozilla-common\_voice\_15-23 & open domain  & read        & monolog \\
        \midrule
        fair-mls-20                 & audiobook     & read        & monolog \\
        \midrule
        pwr-azon\_read-20           & scientific    & read        & monolog \\
        \midrule
        pwr-azon\_spont-20          & scientific    & spontaneous & monolog \\
        \midrule
        pwr-maleset-unk             & commands      & read        & monolog \\
        \midrule
        pwr-shortwords-unk          & commands      & read        & monolog \\
        \midrule
        pwr-viu-unk                 & commands      & read        & monolog \\
        \midrule
        google-fleurs-22            & wikipedia     & read        & monolog \\
        \midrule
        polyai-minds14-21           & banking       & read        & monolog \\
        \bottomrule
    \end{tabular}
\end{table}

\begin{table}[htbp]
    \caption{PELCRA for BIGOS dataset subset domains and speech types.}
    \label{tab:domains_speech_type_pelcra}
    \centering
    \begin{tabular}{c c c c}
        \toprule
        \textbf{Codename} & \textbf{Domain} & \textbf{Speech type} & \textbf{Interaction type} \\
        \midrule
        ul-diabiz\_poleval-22     & customer service & spontaneous & dialog \\
        \midrule
        ul-spokes\_biz\_bio-23    & open domain      & spontaneous & dialog \\
        \midrule
        ul-spokes\_biz\_int-23    & open domain      & spontaneous & dialog \\
        \midrule
        ul-spokes\_biz\_luz-23    & open domain      & spontaneous & dialog \\
        \midrule
        ul-spokes\_biz\_pod-23    & open domain      & spontaneous & dialog \\
        \midrule
        ul-spokes\_biz\_pres-23   & open domain      & spontaneous & dialog \\
        \midrule
        ul-spokes\_biz\_vc-23     & open domain      & spontaneous & dialog \\
        \midrule
        ul-spokes\_biz\_vc2-23    & open domain      & spontaneous & dialog \\
        \midrule
        ul-spokes\_biz\_wyw-23    & open domain      & spontaneous & dialog \\
        \midrule
        ul-spokes\_mix\_emo-18    & open domain      & spontaneous & dialog \\
        \midrule
        ul-spokes\_mix\_luz-18    & open domain      & spontaneous & dialog \\
        \midrule
        ul-spokes\_mix\_parl-18   & open domain      & spontaneous & monolog \\
        \bottomrule
    \end{tabular}
\end{table}
\subsection{Dataset contents details}\label{app:dataset_features_env_device}
Tables \ref{tab:speech_source_env_device_bigos} and \ref{tab:speech_source_env_device_pelcra} present information on sources, acoustic environments and audio recording devices for BIGOS and PELCRA datasets, respectively.

\begin{table}[h]
    \caption{BIGOS dataset subset speakers, environments, and devices.}
    \label{tab:speech_source_env_device_bigos}
    \centering
    \begin{tabular}{c c c c}
        \toprule
        \textbf{Codename} & \textbf{Speech source} & \textbf{Acoustic environment} & \textbf{Audio device} \\
        \midrule
        pjatk-clarin\_studio-15     & volunteers      & quiet   & studio mic \\
        \midrule
        pjatk-clarin\_mobile-15     & volunteers      & quiet   & mobile phone \\
        \midrule
        mailabs-corpus\_librivox-19 & volunteers      & quiet   & various \\
        \midrule
        mozilla-common\_voice\_15-23 & crowd          & various & various \\
        \midrule
        fair-mls-20                 & volunteers      & various & various \\
        \midrule
        pwr-azon\_read-20           & volunteers      & quiet   & studio mic \\
        \midrule
        pwr-azon\_spont-20          & public speakers & mixed   & lavalier \\
        \midrule
        pwr-maleset-unk             & volunteers      & quiet   & studio mic \\
        \midrule
        pwr-shortwords-unk          & volunteers      & quiet   & studio mic \\
        \midrule
        pwr-viu-unk                 & volunteers      & quiet   & studio mic \\
        \midrule
        google-fleurs-22            & volunteers      & quiet   & mobile phone \\
        \midrule
        polyai-minds14-21           & crowd           & quiet   & mobile phone \\
        \bottomrule
    \end{tabular}
\end{table}

\begin{table}[htbp]
    \caption{PELCRA for BIGOS subsets speakers, environments, and devices.}
    \label{tab:speech_source_env_device_pelcra}
    \centering
    \begin{tabular}{c c c c}
        \toprule
        \textbf{Codename} & \textbf{Speech source} & \textbf{Acoustic environment} & \textbf{Audio device} \\
        \midrule
        ul-diabiz\_poleval-22     & volunteers      & quiet   & telephone \\
        \midrule
        ul-spokes\_biz\_bio-23    & volunteers      & quiet   & lavalier mic \\
        \midrule
        ul-spokes\_biz\_int-23    & volunteers      & quiet   & lavalier mic \\
        \midrule
        ul-spokes\_biz\_luz-23    & volunteers      & quiet   & lavalier mic \\
        \midrule
        ul-spokes\_biz\_pod-23    & public speakers & quiet   & various \\
        \midrule
        ul-spokes\_biz\_pres-23   & public speakers & quiet   & various \\
        \midrule
        ul-spokes\_biz\_vc-23     & volunteers      & quiet   & lavalier mic \\
        \midrule
        ul-spokes\_biz\_vc2-23    & volunteers      & quiet   & lavalier mic \\
        \midrule
        ul-spokes\_biz\_wyw-23    & volunteers      & quiet   & lavalier mic \\
        \midrule
        ul-spokes\_mix\_emo-18    & volunteers      & quiet   & lavalier mic \\
        \midrule
        ul-spokes\_mix\_luz-18    & volunteers      & quiet   & lavalier mic \\
        \bottomrule
    \end{tabular}
\end{table}

\subsection{Audio content size metrics}\label{app:dataset_stats}
Tables \ref{tab:size_metrics_bigos} and \ref{tab:size_metrics_pelcra} present information about number of available transcribed speech material, audio files and recorded speakers for BIGOS and PELCRA datasets, respectively.
\begin{table}[htbp]
    \caption{Audio content size metrics for BIGOS dataset}
    \label{tab:size_metrics_bigos}
    \centering
    \begin{tabular}{l r r r}
        \toprule
        \textbf{Subset} & \textbf{Transcribed audio [h]} & \textbf{Samples} & \textbf{Speakers} \\
        \midrule
        fair-mls-20            & 107.86 & 26072  & 24   \\
        \midrule
        google-fleurs-22       & 12.07  & 3937   & 3    \\
        \midrule
        mailabs-corpus\_librivox-19 & 32.14  & 14862  & 2    \\
        \midrule
        mozilla-common\_voice\_15-23 & 53.00  & 36910  & 2920 \\
        \midrule
        pjatk-clarin\_mobile-15 & 12.48  & 3495   & 117  \\
        \midrule
        pjatk-clarin\_studio-15 & 56.43  & 13810  & 553  \\
        \midrule
        polyai-minds14-21      & 3.07   & 562    & 3    \\
        \midrule
        pwr-azon\_read-20      & 5.72   & 2788   & 29   \\
        \midrule
        pwr-azon\_spont-20     & 2.14   & 456    & 27   \\
        \midrule
        pwr-maleset-unk        & 6.38   & 4738   & 3    \\
        \midrule
        pwr-shortwords-unk     & 1.43   & 939    & 3    \\
        \midrule
        pwr-viu-unk            & 1.04   & 2703   & 3    \\
        \midrule
        total                  & 293.76 & 111272 & 3945 \\
        \bottomrule
    \end{tabular}
\end{table}

\begin{table}[htbp]
    \caption{Audio content size metrics for another dataset}
    \label{tab:size_metrics_pelcra}
    \centering
    \begin{tabular}{l r r r}
        \toprule
        \textbf{Subset} & \textbf{Transcribed audio [h]} & \textbf{Samples} & \textbf{Speakers} \\
        \midrule
        ul-diabiz\_poleval-22  & 9.83   & 8950   & 170  \\
        \midrule
        ul-spokes\_biz\_bio-23 & 137.98 & 54917  & 158  \\
        \midrule
        ul-spokes\_biz\_int-23 & 2.25   & 1109   & 9    \\
        \midrule
        ul-spokes\_biz\_luz-23 & 74.27  & 41966  & 158  \\
        \midrule
        ul-spokes\_biz\_pod-23 & 55.00  & 22807  & 113  \\
        \midrule
        ul-spokes\_biz\_pres-23 & 32.25  & 17174  & 55   \\
        \midrule
        ul-spokes\_biz\_vc-23  & 52.07  & 45272  & 78   \\
        \midrule
        ul-spokes\_biz\_vc2-23 & 81.04  & 25802  & 84   \\
        \midrule
        ul-spokes\_biz\_wyw-23 & 28.21  & 11357  & 38   \\
        \midrule
        ul-spokes\_mix\_emo-18 & 25.61  & 24329  & 40   \\
        \midrule
        ul-spokes\_mix\_luz-18 & 18.74  & 20919  & 21   \\
        \midrule
        ul-spokes\_mix\_parl-18 & 12.27  & 8656   & 48   \\
        \midrule
        total                  & 529.52 & 283258 & 972  \\
        \bottomrule
    \end{tabular}
\end{table}

\newpage
\subsection{Evaluated ASR system details}\label{evaluation_results}
\begin{itemize}

    \item \textbf{Google Cloud Speech-to-Text}\footnote{https://cloud.google.com/speech-to-text} supports more than 125 languages and variants. Google's service offers several useful features, such as noise cancelation, support for streaming, automatic punctuation, and the capability to recognize specific phrases or words when provided with context (e.g., specialized vocabulary or formats for spoken numbers, addresses, years, currencies, etc.). For selected languages, it also provides domain-specific models, multichannel audio support, and filtering of profanity content. Two generations of service are available: v1\footnote{https://cloud.google.com/speech-to-text/docs/speech-to-text-requests?hl=en} and v2.\footnote{https://cloud.google.com/speech-to-text/v2/docs?hl=en} For Polish, multiple model variants are available and were evaluated: \emph{v1\_default}, \emph{v1\_latest\_long}, \emph{v1\_latest\_short}, \emph{v1\_command\_and\_search}, \emph{v2\_long} and \emph{v2\_short}.
    
    \item \textbf{Microsoft's Azure Speech Service} \footnote{https://azure.microsoft.com/en-us/products/cognitive-services/speech-to-text} as of May 2023 supports more than 100 languages and variants. In addition to standard transcription, the Azure Speech Service supports continuous real-time speech recognition and provides robust noise reduction capabilities. It allows users to apply custom models to improve the accuracy of domain-specific terminology. Additional services include text search or analytics on transcribed content, as well as speaker diarization. The \emph{latest default}  model for Polish (dated for January 2023) was used, as no specialized model types support this language.
    
    \item \textbf{Whisper} \footnote{https://github.com/openai/whisper/tree/main} is an ASR system developed by the OpenAI company. It is trained on a large amount of weakly supervised multilingual and multitask data collected from the Internet \cite{whisper22}. According to the literature, Whisper is capable of handling different languages, dialects, and accents, demonstrating strong performance in diverse applications when evaluated on well-known benchmark datasets, e.g. Common Voice \cite{whisper22}. Whisper is available via a web API or as a pre-trained model for local use. Five versions of models of varying sizes are available for free download. The large model is available in 3 versions. 
    
\begin{table}[htbp]
    \caption{Model sizes and availability of English-only and Multilingual models.}
    \label{tab:model_sizes}
    \centering
    \begin{tabular}{l l l l}
        \toprule
        \textbf{Size} & \textbf{Parameters} & \textbf{English-only model} & \textbf{Multilingual model} \\
        \midrule
        tiny   & 39 M   & Yes & Yes \\
        \midrule
        base   & 74 M   & Yes & Yes \\
        \midrule
        small  & 244 M  & Yes & Yes \\
        \midrule
        medium & 769 M  & Yes & Yes \\
        \midrule
        large  & 1550 M & No  & Yes \\
        \bottomrule
    \end{tabular}
\end{table}

source: \href{https://github.com/openai/whisper/blob/main/model-card.md}
For this benchmark, the commercial model available via API and eight locally run models were used.

    \item \textbf{NVIDIA NeMo} is the ASR system based on the \emph{quartznet} model, which consists of 79 layers and has a total of 18.9 million parameters. \cite{Kriman2019QuartzNet:Convolutions}  Three models supporting the Polish language are available: \emph{stt\_pl\_fastconformer\_hybrid\_large\_pc}, \emph{stt\_pl\_quartznet15x5} and \emph{stt\_multilingual\_fastconformer\_hybrid\_large\_pc}. The English version was trained on \~3,000 hours of public English data.  Polish models were fine-tuned from English to Polish on the \emph{Mozilla Common Voice (MCV)} Dataset. \cite{Ardila2019CommonCorpusb}. All models are available for free use under a CC-BY-NC license.

    \item \textbf{MMS}: Facebook AI's massive multilingual pre-trained model for speech ("MMS"). It was pre-trained on about 500,000 hours of speech data in more than 1,400 languages \cite{Pratap2023ScalingLanguages}. The MMS system supports over 1000 languages and other speech processing tasks such as \textit{ Text-to-Speech (TTS)} generation and \textit{Speech Language Identification (LID)} \footnote{https://huggingface.co/spaces/mms-meta/MMS}. The MMS system is available for free\footnote{https://huggingface.co/facebook/mms-1b-all} under the CC-BY-NC 4.0 license. The following versions of the fine-tuned model of ASR are available:
    \begin{itemize}
        \item \emph{1b-fl102}  - 1 billion parameter model fine-tuned on \emph{FLEURS} Dataset \cite{Conneau2022FLEURS:Speech}
        \item \emph{1b-l1107}  - 1 billion parameter model fine-tuned \emph{MMS-lab} \cite{Pratap2023ScalingLanguages} Dataset.
        \item \emph{1b-all}  - 1 billion parameter model fine-tuned on \textit{ MMS-lab, FLEURS, CommonVoice, MLS} and\textit{VoxPopuli} datasets. \cite {Ardila2019CommonCorpusb, Pratap2023ScalingLanguages, Pratap2020MLS:Research, voxpopuli2021} 
    \end{itemize}
    
    \item \textbf{Wav2Vec} is the automated speech recognition (ASR) system created by Facebook AI. It employs self-supervision to learn from unlabeled training data. Upon its launch in 2020, wav2vec2 exceeded the top semi-supervised approach with only a fraction of labeled training data \cite{Hsu2021ROBUSTPRE-TRAINING}. Two models fine-tuned for Polish are available on the Hugging Face platform: \emph{xls-r-1b-polish} and \emph{large\_xlsr-53-polish}.

    \item \textbf{Assembly AI}\footnote{\href{https://www.assemblyai.com/}{Assembly AI}} provides an advanced automatic speech recognition service supporting multiple languages. Key features include real-time transcription, automatic punctuation, and robust noise cancellation. The service supports domain-specific vocabulary through custom models, filtering of sensitive content and integration with various platforms via a web API. The system is designed to handle diverse accents and dialects, ensuring high accuracy across different use cases. According to the authors, their system "leverages a diverse training Dataset comprising unsupervised (12.5M hours), supervised (188k hours), and pseudo-labeled (1.6M hours) data across four languages”\cite{Ramirez2024AnatomyASR}. It is also reported that the \textit{Universal-1} model achieves comparative WER scores to larger and more computationally expensive models, such as Whisper large and Canary-1B.\cite{Ramirez2024AnatomyASR}. The amount of training data for Polish is not reported.

\end{itemize}

\begin{table}[htbp]
    \caption{ASR systems evaluated in the study.}
    \label{tab:asr_systems_evaluated_names}
    \centering
    \begin{tabular}{c c c}
        \toprule
        \textbf{Shortname} & \textbf{System} & \textbf{Model} \\
        \midrule
        assembly\_best         & assembly\_ai & best \\
        \midrule
        assembly\_nano         & assembly\_ai & nano \\
        \midrule
        azure\_latest          & azure        & latest \\
        \midrule
        google\_cmd\_search    & google       & command\_and\_search \\
        \midrule
        google\_default        & google       & default \\
        \midrule
        google\_long           & google       & latest\_long \\
        \midrule
        google\_short          & google       & latest\_short \\
        \midrule
        google\_v2\_long       & google\_v2   & long \\
        \midrule
        google\_v2\_short      & google\_v2   & short \\
        \midrule
        mms\_all               & mms          & 1b-all \\
        \midrule
        mms\_102               & mms          & 1b-fl102 \\
        \midrule
        mms\_1107              & mms          & 1b-l1107 \\
        \midrule
        nemo\_multilang        & nemo         & stt\_multilingual\_fastconformer\_hybrid\_large\_pc \\
        \midrule
        nemo\_pl\_confromer    & nemo         & stt\_pl\_fastconformer\_hybrid\_large\_pc \\
        \midrule
        nemo\_pl\_quartznet    & nemo         & stt\_pl\_quartznet15x5 \\
        \midrule
        w2v-53-pl              & wav2vec2     & large-xlsr-53-polish \\
        \midrule
        w2v-1b-pl              & wav2vec2     & xls-r-1b-polish \\
        \midrule
        whisper\_cloud         & whisper\_cloud & whisper-1 \\
        \midrule
        whisper\_base          & whisper\_local & base \\
        \midrule
        whisper\_large\_v1     & whisper\_local & large-v1 \\
        \midrule
        whisper\_large\_v2     & whisper\_local & large-v2 \\
        \midrule
        whisper\_large\_v3     & whisper\_local & large-v3 \\
        \midrule
        whisper\_medium        & whisper\_local & medium \\
        \midrule
        whisper\_small         & whisper\_local & small \\
        \midrule
        whisper\_tiny          & whisper\_local & tiny \\
        \bottomrule
    \end{tabular}
\end{table}

\begin{table}[htbp]
    \caption{Evaluated ASR systems usage cost and license type.}
    \label{tab:asr_systems_evaluated_cost_license}
    \centering
    \begin{tabular}{c c c}
        \toprule
        \textbf{Shortname}  & \textbf{Usage cost} & \textbf{License} \\
        \midrule
        assembly\_best       & commercial & Proprietary \\
        \midrule
        assembly\_nano       & commercial & Proprietary \\
        \midrule
        azure\_latest        & commercial & Proprietary \\
        \midrule
        google\_cmd\_search  & commercial & Proprietary \\
        \midrule
        google\_default      & commercial & Proprietary \\
        \midrule
        google\_long         & commercial & Proprietary \\
        \midrule
        google\_short        & commercial & Proprietary \\
        \midrule
        google\_v2\_long     & commercial & Proprietary \\
        \midrule
        google\_v2\_short    & commercial & Proprietary \\
        \midrule
        mms\_all             & free       & CC-BY-NC \\
        \midrule
        mms\_102             & free       & CC-BY-NC \\
        \midrule
        mms\_1107            & free       & CC-BY-NC \\
        \midrule
        nemo\_multilang      & free       & CC-BY \\
        \midrule
        nemo\_pl\_confromer  & free       & CC-BY \\
        \midrule
        nemo\_pl\_quartznet  & free       & CC-BY \\
        \midrule
        w2v-53-pl            & free       & Apache \\
        \midrule
        w2v-1b-pl            & free       & Apache \\
        \midrule
        whisper\_cloud       & commercial & Proprietary \\
        \midrule
        whisper\_base        & free       & MIT \\
        \midrule
        whisper\_large\_v1   & free       & MIT \\
        \midrule
        whisper\_large\_v2   & free       & MIT \\
        \midrule
        whisper\_large\_v3   & free       & MIT \\
        \midrule
        whisper\_medium      & free       & MIT \\
        \midrule
        whisper\_small       & free       & MIT \\
        \midrule
        whisper\_tiny        & free       & MIT \\
        \bottomrule
    \end{tabular}
\end{table}

\subsection{Normalization methods}
Table \ref{tab:normalization_methods} contains overview of  scope of normalization of each available method.
\begin{table}[htbp]
    \caption{Methods of normalizing references and hypotheses.}
    \label{tab:normalization_methods}
    \centering
    \begin{tabular}{>{\raggedright}p{3cm}  >{\raggedright\arraybackslash}p{10cm}}
        \toprule
        \textbf{Normalization method} & \textbf{Scope} \\ 
        \midrule
        blanks removal & Elimination of superfluous white spaces. \\ 
        \midrule
        lowercasing & Conversion of all characters to lowercase. \\ 
        \midrule
        punctuation removal & Removal of punctuation symbols. \\ 
        \midrule
        lexicon-based normalization & Removal of specific words e.g. fillers "um", "mhm" etc. Unification of spelling e.g. Kissindżer -> Kissinger \\ 
        \midrule
        tags removal & Removal of tags e.g. 'trunc' in PELCRA Dataset. \\ 
        \bottomrule
    \end{tabular}
\end{table}

\subsection{Evaluation results}\label{app:evaluation_results}

\begin{figure}
    \centering
    \includegraphics[width=0.75\linewidth]{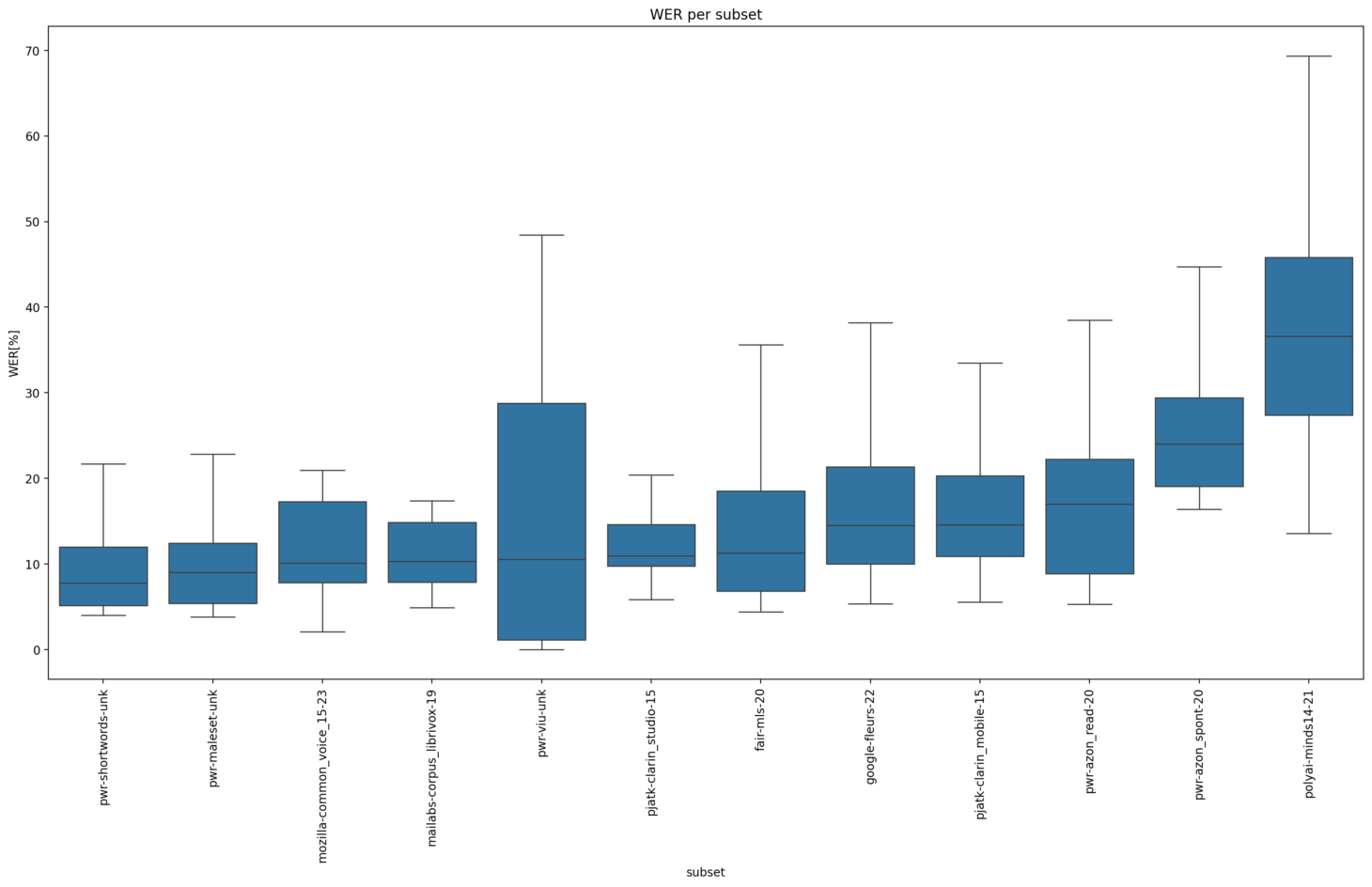}
    \caption{ASR systems accuracy across speaker age groups.}
    \label{fig:wer_age_groups}
\end{figure}

\begin{figure}
    \centering
    \includegraphics[width=1\linewidth]{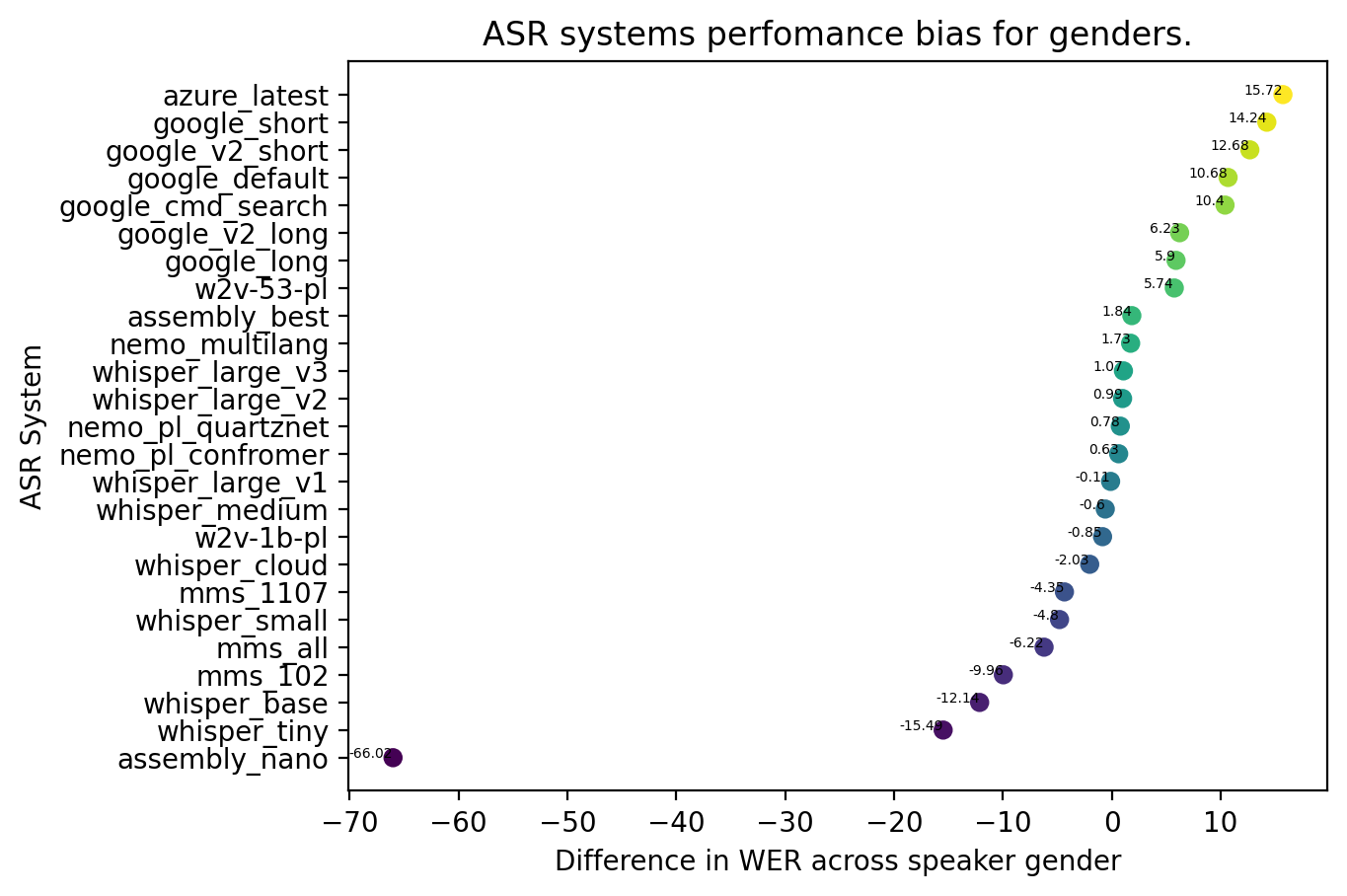}
    \caption{ASR systems accuracy across speaker gender groups.}
    \label{fig:wer_gender}
\end{figure}

\subparagraph{Accuracy per speaker genders}
Figure \ref{fig:wer_gender} shows the difference in WER for the speaker groups of different gender. Positive values indicate bias toward male speakers, while negative values indicate bias toward female speakers. Values close to zero indicate lack of bias.

\begin{figure}
    \centering
    \includegraphics[width=1\linewidth]{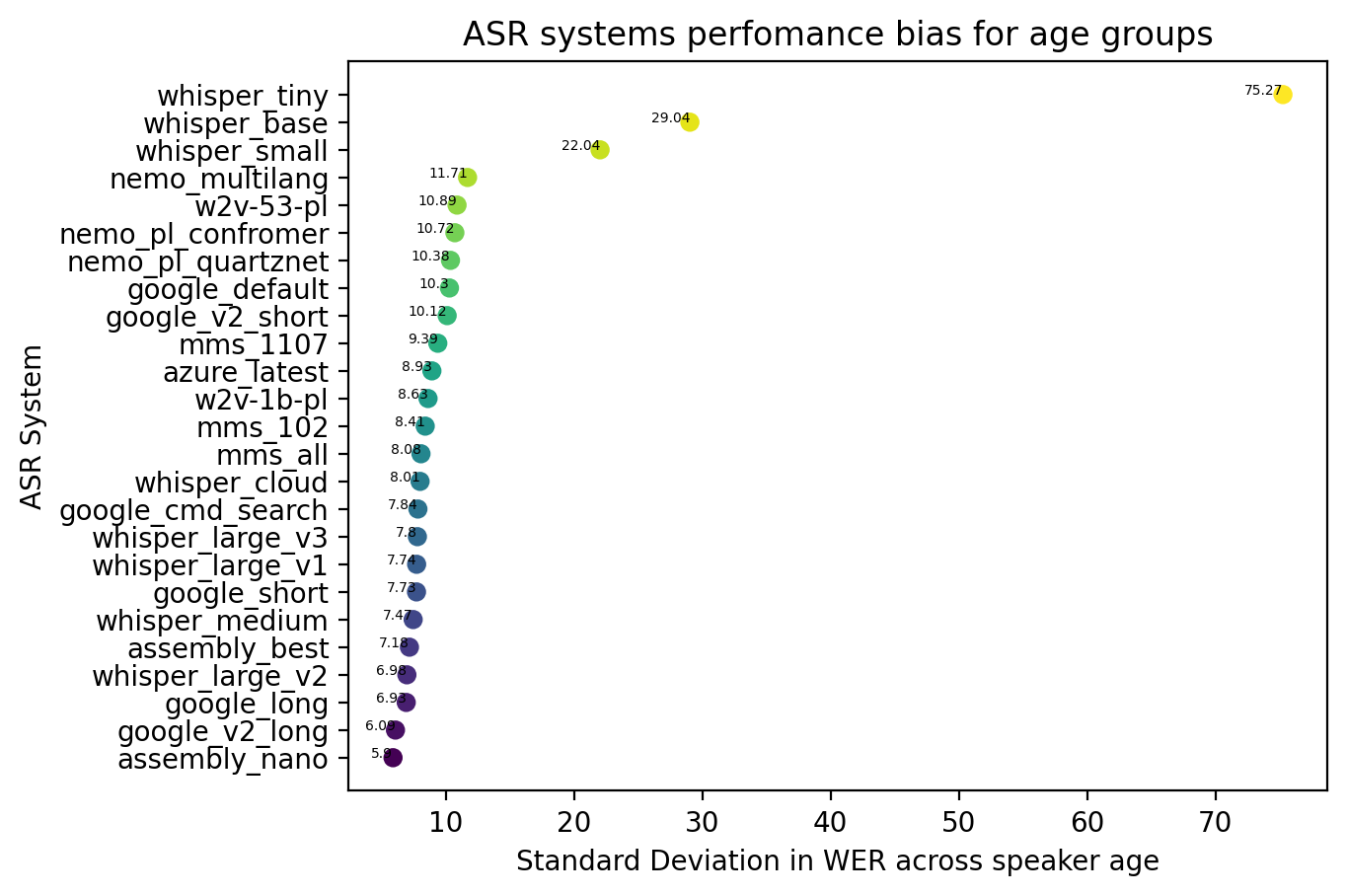}
    \caption{ASR systems accuracy across speaker age groups.}
    \label{fig:wer_age_pelcra}
\end{figure}

\subparagraph{Accuracy per speaker age groups}
Table \ref{tab:age_groups_wer_pelcra} shows the mean WER for age groups in the PELCRA dataset. Figure \ref{fig:wer_age_pelcra} shows the standard deviation of WER in all age groups. Lower values indicate a more consistent accuracy for all groups.
\begin{table}[htbp]
    \caption{Mean WER across systems and age ranges. PELCRA dataset.}
    \label{tab:age_groups_wer_pelcra}
    \centering
    \begin{tabular}{c r r r r r r r}
        \toprule
        \textbf{System} & \textbf{20s} & \textbf{30s} & \textbf{40s} & \textbf{60s} & \textbf{70s} & \textbf{Std.} & \textbf{Range} \\
        \midrule
        assembly\_nano         & 62.88 & 66.14 & 78.59 & 70.49 & 70.46 & 5.9  & 15.71 \\
        \midrule
        google\_v2\_long       & 33.84 & 36.31 & 39.62 & 23.64 & 30.68 & 6.09 & 15.98 \\
        \midrule
        google\_long           & 38.91 & 37.93 & 39.33 & 24.23 & 28.64 & 6.93 & 15.1  \\
        \midrule
        whisper\_large\_v2     & 30.01 & 27.73 & 31.39 & 15.49 & 32.88 & 6.98 & 17.39 \\
        \midrule
        assembly\_best         & 28.37 & 30.24 & 38.3  & 20.94 & 37.66 & 7.18 & 17.36 \\
        \midrule
        whisper\_medium        & 28.73 & 37.38 & 36.84 & 19.16 & 32.74 & 7.47 & 18.22 \\
        \midrule
        google\_short          & 31.8  & 32.13 & 46.57 & 25.47 & 33.74 & 7.73 & 21.1  \\
        \midrule
        whisper\_large\_v1     & 32.65 & 26.38 & 38.59 & 18.24 & 32.77 & 7.74 & 20.35 \\
        \midrule
        whisper\_large\_v3     & 29.13 & 26.69 & 38.05 & 16.87 & 32.13 & 7.8  & 21.18 \\
        \midrule
        google\_cmd\_search    & 39.71 & 46.87 & 53.46 & 32.68 & 40.94 & 7.84 & 20.78 \\
        \midrule
        whisper\_cloud         & 24.32 & 26.5  & 32.41 & 14.13 & 34.5  & 8.01 & 20.37 \\
        \midrule
        mms\_all               & 39.45 & 48.35 & 55.13 & 34.22 & 42.68 & 8.08 & 20.91 \\
        \midrule
        mms\_102               & 50.05 & 53.28 & 62.99 & 43.78 & 63.16 & 8.41 & 19.38 \\
        \midrule
        w2v-1b-pl              & 49.6  & 50.59 & 60.09 & 36.41 & 45.07 & 8.63 & 23.68 \\
        \midrule
        azure\_latest          & 44.67 & 44.92 & 43.38 & 29.78 & 26.6  & 8.93 & 18.32 \\
        \midrule
        mms\_1107              & 52.41 & 50.58 & 61.26 & 35.3  & 47.75 & 9.39 & 25.96 \\
        \midrule
        google\_v2\_short      & 36.64 & 42.1  & 53.69 & 28.22 & 30.95 & 10.12 & 25.47 \\
        \midrule
        google\_default        & 47.32 & 51.8  & 58.04 & 31.94 & 39.17 & 10.3  & 26.1  \\
        \midrule
        nemo\_pl\_quartznet    & 44.83 & 50.18 & 62.07 & 33.55 & 51.13 & 10.38 & 28.52 \\
        \midrule
        nemo\_pl\_confromer    & 44.68 & 52.52 & 62.65 & 34.39 & 54.62 & 10.72 & 28.26 \\
        \midrule
        w2v-53-pl              & 57.59 & 62.35 & 69.96 & 41.41 & 50.99 & 10.89 & 28.55 \\
        \midrule
        nemo\_multilang        & 53.25 & 60.79 & 64.07 & 33.88 & 52.67 & 11.71 & 30.19 \\
        \midrule
        whisper\_small         & 36.7  & 47.93 & 82.85 & 24.1  & 41.72 & 22.04 & 58.75 \\
        \midrule
        whisper\_base          & 55.58 & 53.59 & 68.69 & 120.52 & 51.49 & 29.04 & 69.03 \\
        \midrule
        whisper\_tiny          & 75.62 & 228.49 & 94.38 & 41.07 & 165.15 & 75.27 & 187.42 \\
        \midrule
        median                 & 39.71 & 47.93 & 55.13 & 31.94 & 40.94  & 8.41  & 21.1 \\
        \midrule
        average                & 42.75 & 51.67 & 54.9  & 34.00 & 46.81  & 12.54 & 31.76 \\
        \midrule
        std                    & 12.43 & 38.63 & 16.52 & 21.53 & 27.07  & 14.01 & 34.76 \\
        \bottomrule
    \end{tabular}
\end{table}

\end{document}